\documentclass[manuscript]{aastex}

\usepackage{ulem}
\usepackage[usenames]{color}

\shorttitle{Modeling hard $\gamma$-ray spectra blazars}
\shortauthors{Zacharopoulou et al.}

\begin{document}

\title{Modeling the hard TeV spectra of  blazars 1ES\,0229+200
and 3C\,66A with an internal absorption scenario}

\author{O. Zacharopoulou\altaffilmark{1},  D. Khangulyan\altaffilmark{2}, F.A. Aharonian\altaffilmark{3,1}, L.  Costamante\altaffilmark{4}}
\affil{$^1$ Max-Planck-Institut f{\"u}r Kernphysik,
Saupfercheckweg 1,
69117 Heidelberg,
Germany}
\affil{$^2$ Institute of Space and Astronautical Science/JAXA, 3-1-1 Yoshinodai, Chuo-ku, Sagamihara, Kanagawa 252-5210, JAPAN} 
\affil{$^3$ Dublin Institute For Advanced Studies, 31 Fitzwilliam Place, Dublin 2, Ireland} 
\affil{$^4$ W.W. Hansen Experimental Physics Laboratory \& Kavli Institute 
for Particle Astrophysics and Cosmology, Stanford University, Stanford, CA 94305, USA}
\email{Olga.Zacharopoulou@mpi-hd.mpg.de}

\begin{abstract}
  We study the applicability of the idea of internal absorption of
  $\gamma$-rays produced through synchrotron radiation of
  ultrarelativistic protons in highly magnetized blobs to
  1ES\,0229+200 and 3C\,66A, the two TeV blazars which show unusually
  hard intrinsic $\gamma$-ray spectra after being corrected for the
  intergalactic absorption.  We show that for certain combinations of
  reasonable model parameters, even with quite modest energy
  requirements, the scenario allows a self-consistent explanation of
  the non-thermal emission of these objects in the keV, GeV, and TeV
  energy bands.

\end{abstract}


\section{Introduction}\label{sec:intro}
Over the last few years  a number of  active galactic nuclei (AGN) 
with redshifts  $z \geq 0.1$ have been detected
in the very high energy (VHE; $E \geq 100$~GeV) 
regime\footnote{See {\it http://tevcat.uchicago.edu/} for an updated list of VHE $\gamma$-ray sources}. 
The detection of VHE $\gamma$-rays from such distant objects implies
serious constraints on the intensity and spectral shape of the
extragalactic background light (EBL).  Traveling over cosmological
distances, high energy $\gamma$-rays are significantly absorbed due to
effective interactions with photons of the EBL
\citep{nikishov62,gould67}.  The level of attenuation depends strongly
on the intensity, spectral shape and redshift-dependence of the EBL.
While robust EBL lower limits can be obtained from galaxy counts
\citep{madau00}, the derivation of the EBL properties based on direct
measurements is quite difficult because of dominant foregrounds
\citep[see for a review][]{hauser01,hauser98}.  In this regard, the
theoretical modeling of the processes which
generate
the EBL \citep[see
e.g.][]{primack08,kneiske10,franceschini08,dominguez10} is an
important aspect of the activity in EBL
studies.

The mean free path of $\gamma$-rays due to interactions with EBL
strongly depends on energy. Therefore the intergalactic absorption 
leads not only to attenuation of the absolute  fluxes, but also to significant changes
in the spectral shape of $\gamma$-rays. 
The proper understanding of this spectral deformation 
is crucial for the correct interpretation of the VHE data from distant AGN. 
It is important to note that because of strong Doppler boosting  
of the non-thermal emission ($F_\gamma \propto \delta^4$)  the $\gamma$-ray 
emission  from the brightest blazars can remain   detectable 
even after  severe intergalactic absorption. 
 
Thus the mere attenuation of the $\gamma$-ray emission 
is not enough to derive robust constraints on the EBL models.
The distortion of the initial  spectral shape of $\gamma$-rays contains more information.   
Since  in the effective absorption regime  the optical depth 
$\tau \geq 1$, even a slight change of the EBL intensity can lead to a strong change of the 
energy-dependent spectral deformation factor  $exp[-\tau(E)]$. 
This allows  quite meaningful upper limits on the EBL in the relevant energy bands, based on the 
condition that the intrinsic spectrum of $\gamma$-rays should have a decent 
form, e.g. be not much harder than $E^{-2}$. On the other hand, the   
absorption-corrected  VHE spectra  of some AGN with $z \geq 0.1$ 
in some cases appear  very  hard, even for 
very a low EBL flux,  with a power-law photon index $\Gamma_{\rm int}$
($dN/dE\equiv N_0 E^{-\Gamma_{\rm int}}$)  quite close to
the hardest conventional value of $\Gamma_{\rm int}=1.5$ \citep{aharonian06*b,aharonian07,franceschini08}. 
In the case of  slightly higher  fluxes of the EBL, the reconstructed spectra  
would get even harder, with $\Gamma_{\rm int}<1.5$. 

Although currently there is a  general consensus in the community that 
the EBL intensity should be quite close to the robust lower limits derived from  galaxy counts,
the possibility of slightly higher  fluxes of the EBL cannot yet be  excluded.  
In particular, using Spitzer data and a profile fitting of the faint fringes 
of galaxies, \citet{levenson08} claimed  a new fiducial value for the contribution 
of galaxies to the EBL at $3.6\,\rm
\mu m$ of $9.0^{+1.7}_ {-0.9}\,\rm nW m^{-2} sr^{-1}$, which exceeds
by a factor of $\sim1.6$ the  flux of the EBL 
suggested by \citet{franceschini08}. Following \citet{levenson08},
\citet{krennrich08}  indicated that for this flux of EBL
the initial (absorption corrected) VHE spectra of distant blazars 1ES~0229+200, 1ES~1218+30.4 and 1ES~1101-232
(located at redshifts $z=0.1396$,\, $0.182$ and $0.186$,
respectively) would have a photon index $\lesssim 1.3$. 
This result would challenge the conventional models for
VHE production in AGN. 

Generally, the X- and $\gamma$-ray non-thermal emission of blazars is
interpreted as a sum of synchrotron and inverse Compton (IC)
components of radiation from relativistic electrons, in the framework
of the so-called synchrotron self-Compton (SSC) or external Compton
(EC) scenarios.  In the case of radiatively efficient models,
i.e. assuming a {\it radiatively cooled} particle distribution, the IC
spectrum in the Thomson limit is expected to be steeper than the
power-law distribution with photon index $1.5$.  This limit does not
depend on the electron initial (injection) spectrum and can be
achieved, for example, in the case of a mono-energetic injection. At
higher energies, the $\gamma$-ray spectrum becomes steeper due to the
Klein-Nishina effect. We note however, that typically the spectra
obtained in the frameworks of SSC scenario are steeper, with photon
indices $\sim2$. Therefore, the spectrum with photon index
$\Gamma_{\rm int}=1.5$ is often referred to as the hardest spectrum
allowed by standard blazar models. However, in the expense of
radiation efficiency it is possible to produce harder VHE spectra
still within the SSC framework, for example assuming a high
lower-energy cutoff in the electron spectrum \citep{katarzynski06}.
The postulation of such a cutoff in the electron spectrum implies very
low efficiency of radiative cooling which, in turn, increases the
requirements for the energy in accelerated electrons and at the same
time requires very small magnetic fields.  Thus, in such scenarios we
face a significant (by orders of magnitude) deviation from
equipartition, $W_{\rm e} >> W_{\rm B}$ \citep[see e.g.][]{tavecchio09}.
 
Alternatively, \citet{aharonian08} have suggested a scenario for the
formation of VHE spectra of almost {\it arbitrary} hardness by
involving additional absorption of VHE $\gamma$-rays interacting with
dense radiation fields in the vicinity of the $\gamma$-ray production
region. The key element in this scenario is the presence of a dense
photon field with a narrow energy distribution or with a sharp low
energy cut-off around $>10\,\rm eV$.  In this case, $\gamma$-rays are
attenuated more effectively at energies $\sim 100\, \rm GeV$ than at
energies $\sim \rm 1-10\,TeV$, and therefore, for large optical depths
($\tau\ge1$), the emerging spectrum in the VHE band should gradually
harden towards higher energies \citep[for detail, see][]{aharonian08}.

While the absorption of high  energy $\gamma$-rays  
in the inner parts of AGN jets is generally  possible, or even unavoidable in some cases
\citep{mcbreen79,liu06,reimer07,sitarek08,liu08,bai09,tavecchio09*b},
the detailed modeling of this process  requires additional assumptions 
concerning the presence of low-frequency radiation fields, 
the location and size of the $\gamma$-ray production region, the 
Doppler factor of the jet, {\it etc.}   We note that currently there is no 
observational evidence excluding the photon field properties required by \citet{aharonian08}, 
also in the case of BL Lacs.  Remarkably, the 
internal absorption hypothesis  provides an alternative explanation for the 
non-thermal X-ray emission, namely as synchrotron radiation of 
secondary (pair-produced)  electrons \citep{aharonian08},  which suggests  a
possible solution to  the problem of low acceleration efficiency in
leptonic models of high energy emission of blazars \citep{costamante09}.

In the original paper, \citet{aharonian08} presented a general
description of the scenario with calculations of model SEDs, but the
obtained spectra were not compared with available data.  In the
present paper, we discuss the multiwavelength properties of the
radiation in the internal absorption scenario, and apply the model to
the data of two distant AGN, namely 1ES~0229+200 ($z=0.1396$) and
3C~66A (estimated at $z=0.444$), detected in TeV band
\citep{aharonian07,aliu09,acciari09*c,reyes09}.  Here we adopt the
proton synchrotron radiation as the source of primary $\gamma$-rays,
and consider the absorption due to $\gamma$-$\gamma$ pair production
both in the $\gamma$-ray production region and in the
surroundings. The synchrotron radiation of secondary pairs gives rise
to an additional lower energy non-thermal component. The latter can be
calculated self-consistently and depends on the primary $\gamma$-ray
spectrum, the target photon field and the relativistic motion of the
$\gamma$-ray production region.

\section{Model description}\label{sec:model}

\begin{figure}
\centering
  \includegraphics[width=0.45\textwidth]{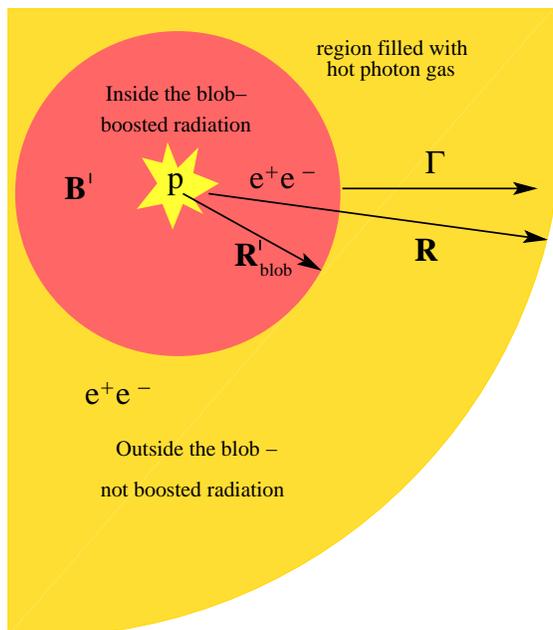} \\
  \caption{A sketch of the model: a blob of  proper radius  
    $R_{\rm blob}'$ (region filled with red color) moves with a bulk
    Lorentz factor $\Gamma$ through a region of typical size $R$
    filled with a hot photon field (yellow region). Protons are
    accelerated and emit synchrotron radiation inside the blob with magnetic field
    strength $B'$. The produced synchrotron emission is
    assumed to be isotropic in the blob frame. The proton-synchrotron
    $\gamma$-rays can be absorbed due to pair production on the soft
    photon field. The pairs created in the blob produce detectable (Doppler boosted)
    synchrotron emission, while the emission of secondary pairs
    produced outside the blob is not Doppler boosted  and therefore not detectable}.
 \label{fig:model}
\end{figure}
A sketch of the model adopted in this paper  is shown in Fig.~\ref{fig:model}, 
and the main ingredients of the model are described in the figure caption.  

\subsection{Primary $\gamma$-rays}
The primary $\gamma$-rays are produced through synchrotron radiation
of protons.  Generally, in such a scenario the energy is stored in the
magnetic field and episodically can be transferred to protons of
extremely high energies forming a non-thermal population of particles
\citep[see for details][]{aharonian00}.  The energy released in
non-thermal protons can be expressed through the strength of the
magnetic field $B'$ and the source radius $R_{\rm blob}'$:
\begin{equation}
E_{\rm tot}'\simeq2\cdot10^{45} \left({\kappa\over 10^{-3}}\right)\left({R_{\rm blob}'\over 10^{15}\,\rm cm}\right)^3\left({B'\over 100\,\rm G}\right)^2 \,\rm erg\,,
\label{eq:energy_total}
\end{equation}
where $\kappa$ is the fraction of the blob magnetic energy transferred
to accelerated  protons (all physical quantities in the blob rest frame are primed).

The proton synchrotron model for blazars works in the case of 
extremely effective acceleration of protons  with an energy  spectrum which 
continues  up to ultra high energies, 
being limited either by the confinement in the accelerator ({so-called ``Hillas criterion''}):
\begin{equation}
E_{\rm Hillas}'\lesssim3\cdot10^{7}\left({R_{\rm blob}'\over 10^{15}\,\rm cm}\right)\left({B'\over 100\, \rm G}\right)\, \rm TeV\,,
\label{eq:max_hillas}
\end{equation}
or by synchrotron losses:
\begin{equation}
E_{\rm max}'\lesssim2\cdot10^{7}\left({B'\over 100\, \rm G}\right)^{-1/2}\, \rm TeV\,.
\label{eq:max_synchrotron}
\end{equation}
The synchrotron cooling time of protons
\begin{equation}
t_{\rm syn}'\simeq5\cdot10^4\left({E'\over 10^{7}\, \rm TeV}\right)^{-1}\left({B'\over 100\,\rm G}\right)^{-2}\, \rm s
\label{eq:cooling}
\end{equation}
is comparable to the proton confinement time assuming Bohm-type diffusion
\begin{equation}
t_{\rm con}'\simeq{3\cdot10^5\over \kappa_{\rm D}}\left({E'\over 10^{7}\, \rm TeV}\right)^{-1}\left({B'\over 100\, \rm G}\right)\left({R_{\rm blob}'\over 10^{15}\, \rm cm}\right)^2\, \rm s \,,
\label{eq:diff}
\end{equation}
where $\kappa_{\rm D}$ is the ratio of the proton diffusion
coefficient to the Bohm one. We note that the confinement time $t_{\rm
  con}'$ cannot be shorter than light crossing time
\begin{equation}
t_{\rm cross}'\simeq3\cdot10^4 \left({R_{\rm blob}'\over 10^{15}\, \rm cm}\right) \, \rm s\,.
\label{eq:light_crossing}
\end{equation}
Given the identical dependencies  of times described by Eqs.(\ref{eq:cooling})-(\ref{eq:diff}) on proton  energy, the cooling regime is defined by
the following parameter:
\begin{equation}
\xi=\frac{t_{\rm con}'}{t_{\rm syn}'}={6\over \kappa_{\rm D}}\left({B'\over 100\, \rm G}\right)^3\left({R_{\rm blob}'\over 10^{15}\, \rm cm}\right)^2\,,
\label{eq:cool_regim}
\end{equation}
implying {\it fast cooling} for $\xi>1$ and {\it slow cooling} for
$\xi<1$. In the case of slow cooling only a fraction $\xi$ of the
proton non-thermal energy will be emitted through the synchrotron
channel. 

In the blob frame the synchrotron emission is expected  to be
isotropic with spectral energy distribution (SED)  extending up to the energy
\begin{equation}
E_{\rm maxH}'\simeq0.2\xi \,\rm TeV
\label{eq:syn_hillas_maximum}
\end{equation}
if the proton maximum energy is given by Eq.() the Hillas criterion
Eq.(\ref{eq:max_hillas}), i.e. $\xi<3$.  In the case of large production region (i.e. $\xi\gg1$), the location of the  SED  maximum is determined by synchrotron losses and is expected to occur at
\begin{equation}
E_{\rm maxS}'\simeq0.4  \, \rm TeV\,.
\label{eq:syn_self_maximum}
\end{equation}

In the laboratory frame the proton emission of such a blob is
characterized by luminosity of
\begin{equation}
L_\gamma\simeq3\cdot10^{46} \left({\kappa\over 10^{-3}}\right)\left({R_{\rm blob}'\over 10^{15}\,\rm cm}\right)^3\left({B'\over 100\,\rm G}\right)^4\left({E'\over 10^7\,\rm TeV}\right)\left({\delta\over 30}\right)^4\, \rm erg\, s^{-1}\,,
\label{eq:energy_total_gamma}
\end{equation}
where $\delta$ is Doppler boosting factor, with typical variability time-scale of 
\begin{equation}
t_{\rm var}\simeq2\cdot10^3\min(1,\xi) \left({\delta\over 30}\right)^{-1}\left({E'\over 10^{7}\,\rm TeV}\right)^{-1}\left({B'\over 100\,\rm G}\right)^{-2}\,\rm s\,.
\label{eq:variability}
\end{equation}
Formally, the VHE spectrum of the boosted proton synchrotron may extend up to
\begin{equation}
E_{\rm max}\simeq10\min(1,\xi/3)\left({\delta\over 30}\right) \,\rm TeV\,.
\label{eq:max_en}
\end{equation}
However, we note that the actual shape of the spectrum close to the cutoff
may be rather smooth, with a significant  fraction of particles above the
formal cutoff energy \citep[e.g. for non-relativistic diffusive shock acceleration; see][]{zirakashvili07}. 
This effect may significantly relax the
constraints imposed by Eq.(\ref{eq:max_en}), given the quadratic dependence
of the synchrotron photon energy on the energy of the parent particle.

For model calculations, in this paper we assume the spectrum of
  non-thermal protons to be a power law with exponential cutoff
  ($N(E)=N_0\,E^{-p}\,exp(-E/E_c)$), with the cutoff energy $E_c$
defined according to Eq.(\ref{eq:max_synchrotron}), i.e. we assume a
very high acceleration efficiency. We consider two cases for the
proton power-law index $p$: (i) the conventional value close to $2$,
and(ii) a very hard case with index $p=-0.5$, as predicted by the {\it
  converter mechanism} \citep{derishev03}. The emission is assumed to
be produced in the {\it slow cooling} regime.  This approximation is
valid for time intervals shorter than the variability time scale
defined by Eq.(\ref{eq:variability}). Under this assumption, the VHE
emission component is characterized by a photon index $1.5$ in the
case (i); and by the hardest possible photon index for the
  synchrotron radiation --namely $2/3$-- in the case of the
converter mechanism (ii). We note that the latter case involves
  VHE spectra harder than conventionally accepted, but this case can
  be realized only in the {\it slow cooling} regime. Otherwise
  (i.e. in the {\it fast cooling} regime), the cooling
  mechanism should modify the proton distribution resulting in VHE spectra
  with photon index close to $1.5$ for dominant synchrotron cooling; or close to $1$ for dominant  adiabatic losses.

\subsection{Internal absorption}
There are several possible sources of UV and soft X-ray emission close
to the base of the jet relate e.g. to accretion disk or corona. This
emission may be reprocessed by matter surrounding the jet. This leads
to the formation of the so-called Broad Line Regions (BLRs), which are
characterized, in the case of powerful blazars, by a size of
$\sim10^{18}$~cm and luminosities of $10^{45}\rm \,erg\,s^{-1}$ \citep[see
e.g.][]{tavecchio08}. Such dense photon fields imply significant
$\gamma\gamma$ absorption, at least if the production region is
located close to the jet base.  The $\gamma\gamma$ optical depth is
estimated as:
\begin{equation}
\tau(E_\gamma)\simeq 0.2\sigma_{\rm T}Rn_{\rm ph}\left(3.5m^2c^4/E_\gamma\right)\,,
\label{eq:depth}
\end{equation}
where $\sigma_{\rm T}$ is the Thomson cross-section, $R$ is the $\gamma$-ray
travel distance in the photon field and $n_{\rm ph}$ is the density of
target photons. Since the size of the region filled  by target-photons  is larger 
than the travel distance $R$, the lower limit on the luminosity of 
the photon field, for the given optical depth $\tau$, is estimated as:
\begin{equation}
 L_{\rm ph}\gtrsim {4\pi R^2{\epsilon_{\rm ph}n_{\rm ph}c\over 4}}\simeq 10^{42}\tau\left(\frac{E_{\gamma}}{100\,\rm GeV}\right)^{-1}  \left( \frac{R}{10^{17}\,\rm cm}\right) \,\rm erg\,s^{-1}\,,
\label{eq:photon_luminosity}
\end{equation}
where $\tau$ is the maximum opacity which occurs for the $\gamma$-ray of 
energy $E_\gamma$. In general, the photon field required for the
internal absorption scenario has a low luminosity and may be
undetectable (it is not Doppler boosted!). 
In order to get an arbitrary hard spectrum after internal absorption, 
we assume the target photon field to be a gray body, i.e. a diluted Planckian distribution,
characterized by the temperature $T$ and the dilution coefficient $\zeta$.

\subsection{Secondary emission}

The energy of the absorbed $\gamma$-ray is transferred to an
electron-positron pair. Since the internal absorption scenario
requires large optical depths $\tau\ge1$, a significant fraction of
energy given by Eq.(\ref{eq:energy_total})  goes to   secondary 
electrons. The observational appearance of these secondaries depends
strongly on the site of their production. Namely, if the pair is  created
outside the blob, emission of these electrons  will not be boosted 
and thus remain undetectable. On the other hand, if the electrons are produced 
in the blob, they will be  isotropized and emit synchrotron
radiation due to the strong magnetic field in the blob.  This
radiation component  can be detected because of strong Doppler boosting. 

In the blob reference frame, the target photon field is strongly
anisotropic. Thus, the optical depth in the blob depends on the
direction of the $\gamma$-ray with respect to bulk velocity. Since we
assume the blob to be homogeneous, 
we introduce the optical depth $\tau_{\rm in}$, averaged over  
the $\gamma$-ray directions (in the blob rest frame)\footnote{ In our calculations the averaged optical 
depth is defined as ${\rm e}^{-\tau_{\rm in}}=<{\rm e}^{-\tau}>$.}, 
to characterize the absorption in the blob. 
The corresponding values are  shown in Table~\ref{table:parameters}.

It is possible to estimate the optical depth $\tau_{\rm in}$ in the blob since
basically all the emission is focused towards the direction of the
proper motion. Indeed, the optical depth for a $\gamma$-ray propagating
in the direction of the proper motion will be
\begin{equation}
\tau_{\rm in}(E_\gamma)\simeq 0.2\sigma_{\rm T}R'_{\rm blob}\Gamma n_{\rm ph}\left(3.5m^2c^4/E_\gamma\right)\,,
\label{eq:bobl_depth}
\end{equation}
where $\Gamma$ is blob bulk Lorentz factor \citep[for detail see e.g.][]{begelman08}. Thus, a simple relation\footnote{This relation was used to derive the sizes of the blob listed in Table~\ref{table:parameters}.}
between size of the BLR region, size of the blob, maximum optical
depth $\tau$, blob optical depth $\tau_{\rm in}$ and bulk Lorentz
factor can be written as
\begin{equation}
{\tau_{\rm in}\over \tau}\simeq {R'_{\rm blob}\Gamma \over R}\,.
\label{eq:depths_ratio}
\end{equation}
This ratio indicates that in the case of a compact region filled with 
photon gas, $R\sim 10^{17}$~cm, the optical depth in the blob is quite
high:
\begin{equation}
\tau_{\rm in}\simeq 0.2\tau \left({R'_{\rm blob}\over 10^{15}\,\rm cm}\right)\left({\Gamma \over 20 }\right)\left({R\over 10^{17}\,\rm cm}\right)^{-1}\, .
\label{eq:in_depth_estimate}
\end{equation}

The injection spectrum of secondary electrons depends on the photon
index of primary $\gamma$-rays, target photon field, bulk Lorentz
factor of the blob and the internal optical depth. If the target
photon field is characterized by a peak energy $\varepsilon$, then 
the maximum injection rate in the
blob occurs at energy 
\begin{equation}
E_{\rm e}'\simeq 5 \left({\varepsilon\over10\,\rm eV}\right)^{-1}\left({\Gamma\over 20}\right)^{-1}\,\rm GeV\,.
\label{eq:inj_en}
\end{equation}
However, we have to note that depending on the slope of the primary $\gamma$-ray
spectrum, this value can change significantly. Since the
synchrotron cooling time of these electrons,
\begin{equation}
t_{\rm syn}'\simeq 40 \left({E_{\rm e}'\over 1\,\rm GeV}\right)^{-1}\left({B'\over 100\,\rm G}\right)^{-2}\,\rm s\,,
\label{eq:electron_colling_time}
\end{equation}
is very short  (compared to  both the typical time
scales for the system and  the Compton cooling time of electrons), 
\begin{equation}
t_{\rm ic}'\simeq 7\cdot10^3 \left({E_{\rm e}'\over 1\,\rm GeV}\right)^{-1}\left({\Gamma\over 20}\right)^{-1}\left({R_{\rm blob}'\over 10^{15}\,\rm cm}\right)\left({\varepsilon\over 10\,\rm eV}\right)^{-1}\left({\tau_{\rm in}\over 1}\right)^{-1}\rm s\,,
\label{eq:electron_ic_time}
\end{equation}
the entire absorbed energy will be immediately released by secondary electrons 
through the synchrotron channel.

In the case of large internal absorption or
high bulk Lorentz factor, the secondary synchrotron component has a
broad distribution centered at  
\begin{equation}
\epsilon_{\rm sec}\simeq 1.5\left({\Gamma\over 20}\right)^{-2}\left({\delta\over 30}\right)\left({\varepsilon\over 10\,\rm eV}\right)^{-2}\left({B'\over 100\,\rm G}\right)\,\rm keV\, .
\label{eq:secondary_maximum}
\end{equation}

The variability time-scale of the
synchrotron  radiation  of secondary pairs is determined by the change of
the injection, i.e. by the change of primary $\gamma$-ray component.  
In the case of small internal opacity and assuming that 
protons are distributed over the energy interval 
between $1\,\rm GeV$ and $10^7\, \rm TeV$ with  $E^{-2}$-type spectrum, 
the luminosity of the secondary synchrotron radiation is 
estimated as 
\begin{equation}
L_{\rm sec}\simeq 10^{44}\left({\tau_{\rm in}\over 0.25}\right) \left({\kappa\over 10^{-3}}\right)\left({R_{\rm blob}'\over 10^{15}\,\rm cm}\right)^3\left({B'\over 100\,\rm G}\right)^4\left({\delta\over 30}\right)^4 \,\rm erg\,s^{-1}\,.
\label{eq:secondary_fraction}
\end{equation}

In the framework of the discussed model an important relation can be
established between the slopes of the intrinsic $\gamma$-ray spectrum
and the highest energy part of the secondary synchrotron
components. This part of the synchrotron spectrum is produced by
electron-positron pairs which are created significantly above the
threshold of the $\gamma\gamma$ interaction, thus it is possible to
use the asymptotic limit of the cross section. Since one of the
secondary electrons receives almost the all parent $\gamma$-ray
energy, the cross section can be approximated as
\begin{equation}
 \frac{{\rm d}\sigma}{{\rm d}E_e}\propto \frac{\delta\left( E_e-E_{\gamma} \right)}{E_{\gamma}}.
\end{equation}
Then, the spectrum of the secondary pairs, which is determined 
by the intrinsic spectrum of VHE $\gamma$-rays, is:
\begin{equation}
\frac{{\rm d}N_e}{{\rm d}E_e{\rm d}t}\propto c\int {\rm d}E_{\gamma}\frac{{\rm d}\sigma}{{\rm d}E_e}\frac{{\rm d}N_{\gamma}}{{\rm d}E_{\gamma}}=\frac{c}{E_e}\frac{{\rm d}N_{\gamma}}{{\rm d}E_e}.
\end{equation}
In particular, if the intrinsic $\gamma$-ray spectrum is a power law in this energy band, 
with a photon index $s$, then, since the
dominant cooling mechanism is synchrotron radiation, the energy
distribution of the secondary leptons is a power law with the index $s+2$
and the high energy part of the synchrotron spectrum is described by a
power law with  photon index $(s+3)/2$. We note that even for a very hard
intrinsic $\gamma$-ray spectrum of $s\sim 1.5$, the synchrotron emission of
secondary pairs will be characterized by a photon index $\sim
2.25$. Such behavior is expected at  energies 
\begin{equation}
\epsilon \ge 200\left({\Gamma\over 20}\right)^{-2}\left({\delta\over 30}\right)\left({\varepsilon\over 10\,\rm eV}\right)^{-2}\left({B'\over 100\,\rm G}\right)\,\rm keV\,.
\label{eq:xray_limit}
\end{equation}

For numerical calculations, we assumed the blob to be homogeneous. The
pair production kernel, i.e. the energy distribution of secondary
electrons produced by a $\gamma$-ray of a certain energy, was
calculated using anisotropic differential pair production cross
section convolved with boosted Planckian distribution and averaged
over the initial $\gamma$-ray direction. The injection rate of
electrons was calculated by convolving the pair production kernel with
proton synchrotron spectrum multiplied by the factor
$\left(1-\exp(-\tau_{\rm in})\right)$.  The energy distribution of
electrons was calculated using the approximation of continuous losses
accounting for dominant synchrotron losses only. The secondary
synchrotron emission was calculated using the obtained distribution of
electrons. The variability properties of this radiation component are
related to the variability of the intrinsic $\gamma$-rays as well as
to the to the change of their absorption rate.

\begin{figure}
  \centering
  \includegraphics[width=1.0\textwidth]{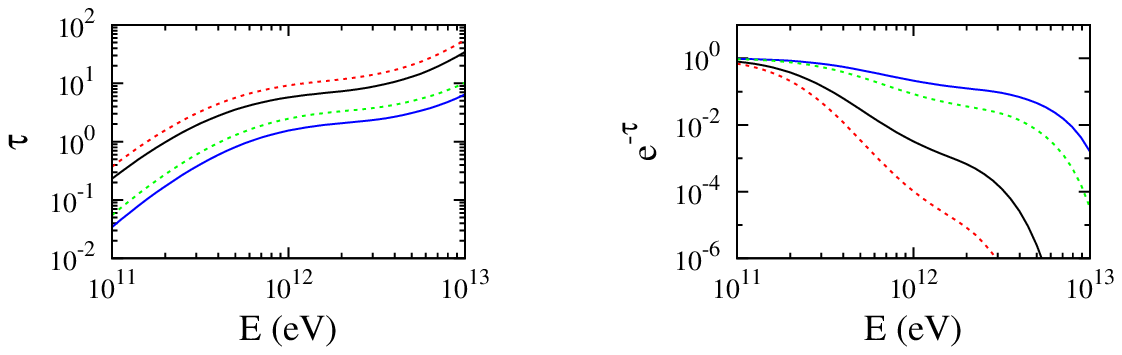} \\
  \caption{{\bf Left panel:} The optical depth $\tau$  for  $\gamma$ rays interacting 
     with the EBL. The upper set of lines is for  $z=0.1396$ (distance of 1ES~0229+200)
     while the lower set is for $z=0.444$ (distance of 3C~66A).  
     The dashed lines are calculated for the EBL model F1.6, the solid lines for the EBL model F1.0.
    {\bf Right panel:} The attenuation factors $e^{-\tau}$ corresponding to the optical depths
     shown in the  left panel.  }
  \label{fig:tau}
\end{figure}

\begin{figure}
  \centering
  \includegraphics[width=0.85\textwidth]{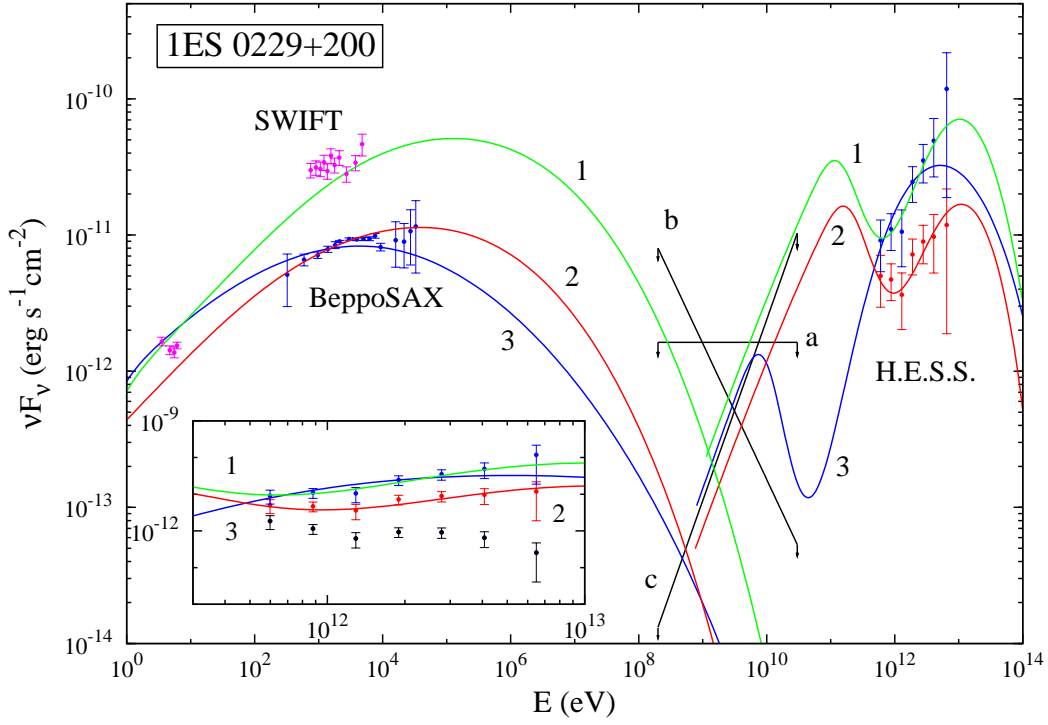} \\
  \caption{ The overall SED of the blazar 1ES~0229+200, together with  model curves calculated 
  with the proton synchrotron scenario plus internal absorption.  
   The set of parameters used in the calculation of 
   Fits~1 (green),~2 (red) and 3 (blue) are presented in  Table~\ref{table:parameters}.
  \textbf{Inner panel:} the VHE spectrum as observed by H.E.S.S. \citep[][black points]{aharonian07}.
   The same spectrum corrected for EBL absorption with model F1.0 (red points) results in    
   an intrinsic power-law index of $\Gamma_{\rm int}\simeq1.5$, while using the higher EBL model 
   F1.6 (blue points) it yields $\Gamma_{\rm int}\simeq1.0$.
  \textbf{Outer panel:}  The same data  as in the inner panel with the addition of the SWIFT data  
   (X-ray and optical bands, magenta points) and of the BeppoSAX data (X-ray band, blue points). 
   The solid black lines between $30\, \rm MeV$ and $30\, \rm GeV$ are the upper limits based
    on FERMI~LAT observations \citep{abdo09}, and calculated assuming  power-law 
    $\gamma$-ray spectra with photon indices  $\Gamma_{\rm int}=2$~(a), $\Gamma_{\rm int}=2.5$~(b) and $\Gamma_{\rm int}=1$~(c). }
  \label{fig:0229}
\end{figure}

\begin{figure}
  \centering
  \includegraphics[width=0.85\textwidth]{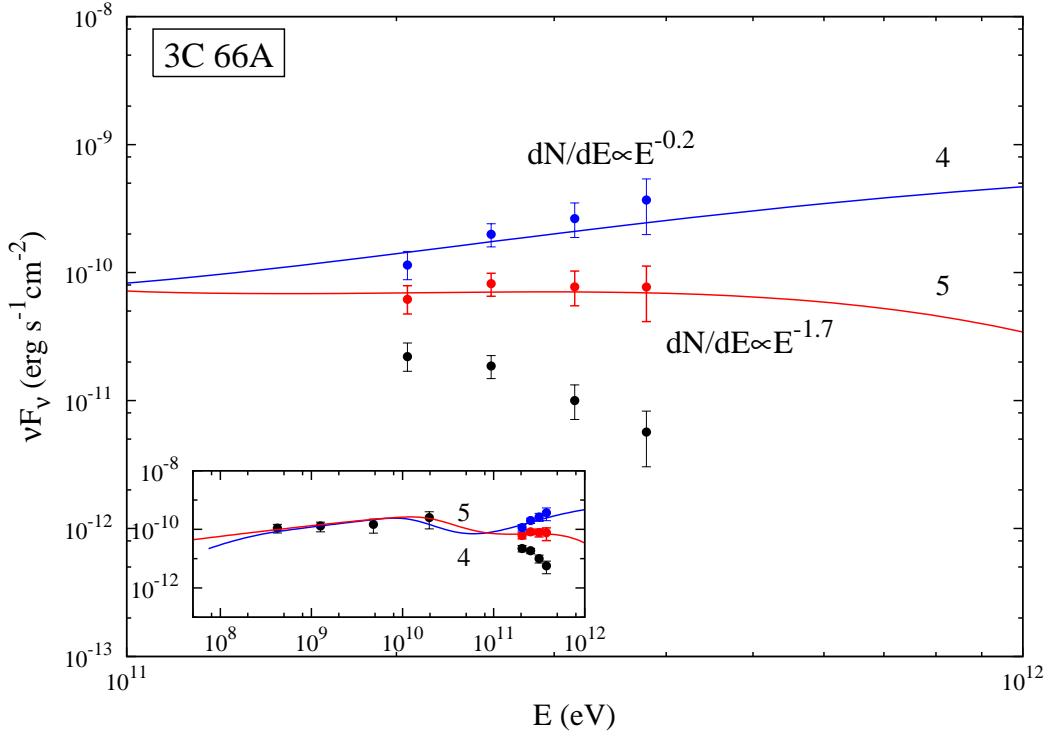} \\
  \caption{ 
  The VERITAS spectrum of 3C~66A \citep[][{\it flare} dataset, black points]{lat66a}.
  The same data corrected for intergalactic absorption  using the EBL model F1.0 (red points) result in
  a power-law intrinsic spectrum ($dN/dE\propto E^{-\Gamma_{\rm int}}$) with photon index 
  $\Gamma_{\rm int}\simeq1.7$. Using the EBL model F1.6 (blue points),
  instead, the photon index becomes $\Gamma_{\rm int}\simeq0.2$. 
The two model lines (labeled 4 and 5) are calculated using the parameters presented 
  in Table~\ref{table:parameters} (Fit 4 and 5).
  \textbf{Inner panel:} Zoom out of the plot to include the GeV band.
  The data points correspond to the FERMI~LAT observations performed simultaneously 
  with VERITAS \citep{lat66a}.
  }
  \label{fig:3c66a}
\end{figure}

\begin{figure}
  \centering
  \includegraphics[width=0.85\textwidth]{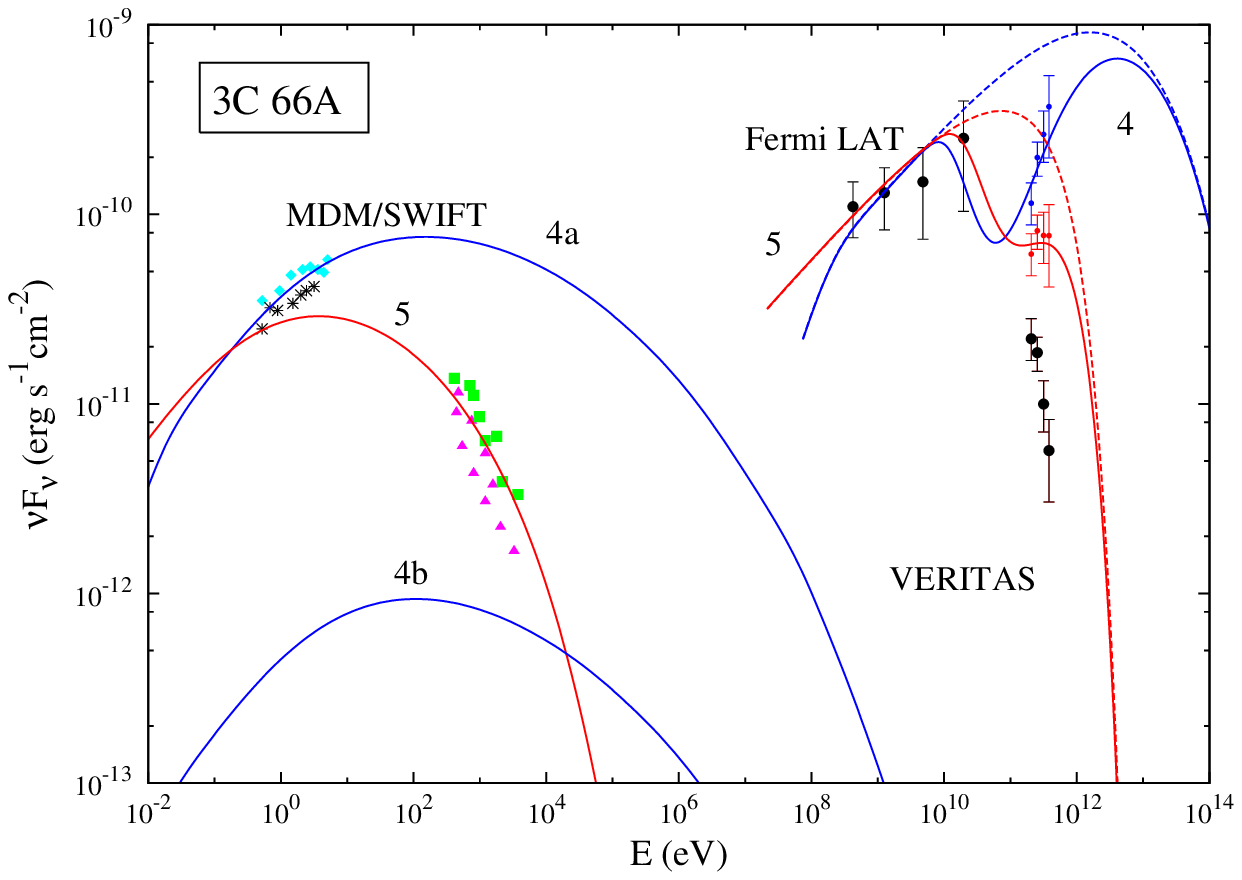} \\
  \caption{ The overall SED of 3C~66A as observed in VHE \citep[VERITAS;][]{lat66a} 
  and  GeV (FERMI~LAT) $\gamma$-rays together with optical (MDM) and X-ray (Swift) 
  data \citep{lat66a}. The curves represent  the model calculations performed for the combinations of parameters 
  reported in Table~\ref{table:parameters}. Fit 4a and 4b differ only in the size of the $\gamma$-ray 
  production region (large or small, respectively), i.e. if the majority of the electron-positron pairs
  are produced inside or outside the relativistically moving blob.
  The dashed lines 4 and 5 show the spectra before internal absorption.}
  \label{fig:3c66amwl}
\end{figure}

\section{Broadband spectra of  1ES~0229+200 and 3C~66A}\label{sec:results}
To demonstrate the potential of the proposed model for the explanation of
very hard intrinsic $\gamma$-ray spectra, we focused on two distant
objects, namely 1ES~0229+200 ($z=0.1396$) and 3C~66A 
\citep[$z=0.444$, though this value is debated, see e.g.][]{lat66a}.
These two BLLacs have different $\gamma$-ray properties.  In
particular, 1ES~0229+200 shows VHE $\gamma$-ray emission without
significant flux or spectral changes between two HESS measurements
separated by one year \citep{aharonian07}.  Moreover, Fermi~LAT was
not able to detect GeV emission from the direction of 1ES~0229+200.
In the SED plot, the upper limit of the GeV flux appears below the TeV
flux corrected for intergalactic absorption.

The blazar 3C~66A shows a variable VHE signal, as seen with VERITAS
\citep{acciari09*c,lat66a}, with a $6\%$ Crab flux flaring episode.
Fermi~LAT collaboration reported a significant GeV $\gamma$-ray excess
from the source.  Moreover, an increase of the GeV flux simultaneously
with the VHE flare was observed \citep{lat66a}.  Importantly, the GeV
flux level exceeds significantly the de-absorbed VHE flux, thus a
smooth connection of these two radiation components with a single
emission seems difficult to achieve, for the assumed redshift z=0.444.

To study the impact of EBL on the VHE spectra, we corrected the
reported $\gamma$-ray spectra for intergalactic absorption using two
versions of the EBL model by \citet{franceschini08}: (i) as in the
original paper (F1.0) and (ii) scaled up by a factor of 1.6 (F1.6).
The latter case was considered in order to satisfy the lower limits
claimed by \citet{levenson08}.  This simple treatment of the EBL and
the related calculations of intergalactic absorption allows us to
ignore many details of different EBL models, and focus on the main
objective of this paper, namely the explanation of hard intrinsic
$\gamma$-ray spectra in blazars. Note that the two EBL
templates used here cover a broad range of different realizations of
the EBL described by recent theoretical or phenomenological models, at
least as long as it concerns the calculated optical depths.

The optical depth for a high energy photon $E_{\gamma}$ traveling
through the intergalactic medium from a source at redshift $z$ to the
observer, taking into account the cosmological distance and the EBL
evolution, is
\begin{equation}
\label{eq:tau_ebl}
\tau_{\gamma\gamma}\left(E_{\gamma},z\right) = c \int\limits_0^z\!{\rm d}z' \frac{{\rm d}l}{{\rm d}z'} \int\limits_0^2 \!{\rm d}x\frac{x}{2} \int\limits_{\frac{2m_e^2c^4}{E_{\gamma}\varepsilon x(1+z')}}^{\infty} \!{\rm d}\varepsilon\, n_{\gamma}(z',\varepsilon)\sigma_{\gamma\gamma}(E_{\gamma}(1+z'),\varepsilon,x)\,,
\end{equation}
where $\frac{dl}{dz'}$ is the cosmological line element;
$x=1-\cos\theta$ is the angle between the interacting photons;
$n_{\gamma}$ is the number density of the EBL as a function of
redshift and soft-photon energy; and $\sigma_{\gamma\gamma}$ is the
pair production cross section.  In Fig.~\ref{fig:tau} the VHE
$\gamma$-ray optical depths (left panel) and attenuation factors
(right panel) for the two blazars are shown, for the two EBL levels:
F1.0 (solid lines) and F1.6 (dashed lines).  The calculated
attenuation was used to reconstruct the initial spectra from the
observed data by H.E.S.S. on 1ES~0229+200 \citep{aharonian07} and by
VERITAS on 3C~66A \citep{lat66a}.  The resulting spectra are
shown in Fig.~\ref{fig:0229} for 1ES~0229+200, and in
Fig.~\ref{fig:3c66a} for 3C~66A.  In both figures black points
correspond to observed data, red points to the spectra reconstructed
with the F1.0 EBL model, and blue points to the spectra reconstructed
with the F1.6 EBL model.

The reconstructed spectra are significantly harder compared to the
observed ones.  In particular, in case of high EBL flux (F1.6),
the spectra have $\Gamma_{\rm int}\simeq 1$ and $0.2$ for 1ES~0229+200 and
3C~66A, respectively; i.e. they would be significantly harder than the
conventional value of $1.5$.

\subsection{The case of 1ES\,0229+200}
To study the case of 1ES~0229+200, we have combined the reconstructed
VHE data with archive X-ray and optical data from SWIFT
\citep{tavecchio09} and BeppoSAX \citep{costamante02}, together with
Fermi~LAT observations \citep{abdo09}.  The observational data are
summarized in Fig.~\ref{fig:0229}.  We have applied the internal
absorption scenario as described in Section~\ref{sec:model} to reproduce
the VHE spectrum together with X-ray spectrum, for both levels of
intergalactic absorption
{and considering both indices  of the power-law proton distribution 
($p\sim2$ and $p=-0.5$).
}

In the case of a soft energy distribution of protons ($p\gtrsim2$), 
{the flux upper limit obtained with Fermi requires a lower-energy   
cutoff ($E'_{\rm le}$) in the proton energy spectrum at very high energies,
with both levels of EBL absorption.}
The exact location of this cutoff depends on different model   
parameters, e.g. intrinsic optical depth, Doppler boosting factor {\it etc.}, 
but to satisfy the Fermi upper limits 
the resulting $\gamma$-ray spectral break  should occur close to $\sim100$~GeV.  
Thus, the cutoff in  the proton spectrum should be located roughly at
\begin{equation}
E'_{\rm le}\simeq2\cdot10^6\left({B'\over100\rm\,G}\right)^{-1/2}\left({\delta\over30}\right)^{-1/2}\rm\,TeV\,.
\end{equation}
This value is  very close to the highest possible energy of the accelerated
protons, thus the Fermi upper limits basically exclude the possibility
of a proton spectrum significantly steeper than ${\rm d}N/{\rm d}E\propto E^{-2}$.

Since the VHE spectrum obtained from 1ES\,0229+200 seems to show no
significant changes on a yearly time scale, i.e. on a time scale much
longer than the one defined by the cooling time (see
Eq. \ref{eq:cooling}), the proton spectrum is expected to be steady.
A steady proton distribution with power-law index $p=2$ can be formed
in two different ways: (i) with an almost mono-energetic continuous
proton injection (e.g. through converter mechanism) in the {\it fast
  synchrotron cooling} regime; and (ii) with a conventional
acceleration spectrum in the {\it slow cooling} regime.  A very hard
steady proton distribution with $p=-0.5$ requires can be formed in the {\it
  slow cooling regime} when an acceleration mechanism similar to
converter mechanism is responsible for the particle acceleration.

The resulting model parameters are summarized  in Table~\ref{table:parameters} 
and the corresponding curves (Fits 1-3) are shown  in Fig.~\ref{fig:0229}.

\subsubsection{F1.0 EBL level}
In the case of the EBL level F1.0, the de-absorbed VHE spectrum has
a photon index close to $\Gamma_{\rm int}\simeq1.5$. 
For a proton distribution with $p=2$, the proton synchrotron radiation
below the peak has a photon index close to $1.5$, i.e.  formally it
can explain the VHE data points without invoking internal absorption.
Thus, in this case the key question is whether the internal absorption
scenario can provide a consistent explanation of the X-ray component.

Given the strict upper limits provided by Fermi,
which are at the level of the extrapolation in the HE band of
the $\Gamma_{\rm int}\simeq1.5$ VHE spectrum,
the available energy budget for the secondary pairs is
quite limited, unless a higher emission can be effectively 
suppressed in the Fermi-LAT band.
This could be achieved either by assuming a broad energy distribution of target photons 
extending to X-ray energies, so to provide a significant attenuation also in the GeV band, 
or by introducing a very high lower-energy cutoff in the proton distribution.
  
In absence of these two conditions, the X-ray synchrotron flux of the secondary pairs 
would be approximately an order of magnitude below the reported X-ray fluxes.
Therefore, in this specific case, the internal absorption scenario
requires additional ad-hoc assumptions 
to provide a self-consistent interpretation of the TeV and X-ray data.

These additional assumptions instead are not needed in the case of a hard
proton spectrum ($p=-0.5$). The latter can provide both the energy budget to explain
the X-ray data and GeV fluxes below the Fermi-LAT limits, as shown 
in Fig.~\ref{fig:0229} (Fit~2, whose corresponding parameters are given 
in Table~\ref{table:parameters}).

\subsubsection{F1.6 EBL level}
In the case of  high EBL flux (model F1.6), the de-absorbed VHE
spectrum has a photon index close to $\Gamma_{\rm int}\simeq1$, i.e. harder than the unabsorbed
synchrotron spectrum from a proton distribution with index $p\sim2$.  
Internal absorption allows the hardening of the TeV spectrum to
the required level, but in the case of a conventional proton
distribution the discrepancy with the Fermi-LAT upper limits is 
very strong.  To avoid the conflict with Fermi-LAT data we need {again} 
to suppress the GeV emission, by introducing additional assumptions 
such as an effective absorption of GeV $\gamma$-rays (e.g. by X-rays) or a
very high lower-energy cutoff (at $10^{6}$~TeV in proton energy).
However, these assumptions can hardly be endorsed without an
additional observational or theoretical justification.

On the other hand, a very hard proton distribution as
predicted by the {\it converter mechanism} \citep{derishev03} can
accommodate the Fermi~LAT upper limits.  Even so, the synchrotron
spectrum would not be sufficiently hard to explain the $\gamma$-ray spectrum
corrected for the higher EBL flux (model F1.6).  In this case the
internal absorption becomes a key requirement to further harden the
initial proton synchrotron spectrum 
(see Fits~1 and 3 in Fig.~\ref{fig:0229}, and the corresponding parameters
in Table~\ref{table:parameters}).

The synchrotron radiation of secondary electron-positron pairs,
calculated self-consistently with the hard VHE component, can explain
the X-ray flux of 1ES~0229+200, with the caveat that the X-ray data
are not simultaneous with the TeV observations. A characteristic
feature of the secondary synchrotron radiation is its broad spectral
extension up to hard X-rays.  This prediction can be tested with hard
X-ray instruments like Suzaku, or with future missions NuStar and
Astro-H.

\subsection{The case of 3C\,66A}
For the blazar 3C~66A, we have combined the VHE spectrum reported by
VERITAS with the spectrum detected by Fermi-LAT
during the VHE flare, together with the available X-ray/optical data
from MDM and Swift \citep{lat66a}.  The observational data are
summarized in Fig.~\ref{fig:3c66a} and \ref{fig:3c66amwl}.  We have
applied the internal absorption scenario as described in
Section~\ref{sec:model} to fit the VHE spectrum, again considering both
levels of EBL absorption (F1.0 and F1.6).
 
To correct for intergalactic absorption, the redshift for the source
we adopted is the one most often cited and used in the literature,
$z=0.444$.  It should be noted, however, that the redshift of this
source is not yet firmly established, and thus one cannot exclude that
the source is located closer. In particular, \citet{prandini10}
suggested that the redshift should not exceed $0.34$.  This conclusion
is based on the belief that the initial VHE $\gamma$-ray spectrum
cannot be harder than the GeV spectrum measured with Fermi~LAT.  On
the other hand, if the redshift is indeed $\gtrsim 0.4$, the TeV and
GeV parts look quite different, and not part of a single
component. Even though, this does not imply that they {must be} of
different origin.  In fact, our model can explain both components with
a single proton population, as parts of the smooth proton synchrotron
spectrum {which is then} deformed by the energy-dependent internal
absorption.

\subsubsection{F1.0 EBL level} 

In this case the de-absorbed TeV spectrum is rather flat, with photon
index $\Gamma_{\rm int}\simeq1.7$, while the HE component is
characterized by a similar photon index $\Gamma\sim1.8$ but with at
higher flux.  A good agreement between the GeV and TeV spectra can be
achieved assuming a proton energy distribution with power-law index
$p=2$.
  
A weak internal absorption (with maximum optical depth of about
$\tau=1.6$) allows modification of the VHE spectrum to the required
photon index (Fit~5 in Fig.~\ref{fig:3c66a}), while the HE part is
reproduced by the unmodified synchrotron spectrum. The synchrotron
emission of secondary pairs can explain the X-ray spectrum obtained
with Swift but not the optical MDM data, which require an additional
radiation component.

The physical parameters used in this model may appear quite extreme
(see Table~\ref{table:parameters}, Fit~5). In particular, the very small
value of the Doppler factor has been chosen to avoid $\gamma$-ray
excess above $1$~TeV, and this consequently leads to a dramatic
increase of the required energy budget. In fact, there is a more
natural way to suppress the flux level above $1$~TeV, namely assuming
a less efficient acceleration process. In this way the Doppler factor
and B-field may be increased, while the required energy budget will be
significantly reduced. The detail study of this possibility will be
discussed elsewhere.

\subsubsection{F1.6 EBL level} 
For the high EBL flux a very small photon index of $\Gamma_{\rm
  int}\simeq0.2$ is required. Remarkably, even such an unusual photon
spectrum can be explained by internal absorption with a higher target
photon temperature and slightly larger optical depth (see
Fig.~\ref{fig:3c66a}, Fit~4).  With a certain combination of model
parameters, the flux of the synchrotron radiation from secondary
electrons can match the levels detected in the optical band, as is
demonstrated in Fig.~\ref{fig:3c66amwl}, Fits~4a and 4b.  For a small production
region, the main fraction of the secondary pairs are produced outside
the blob.  Their radiation is not Doppler boosted and, therefore
cannot be detected.  For the pairs produced inside the blob, the
secondary synchrotron radiation is Doppler boosted and thus it can
contribute significantly to the observed fluxes.  We note, however,
that for this source we did not succeed to find a combination of
parameters which could explain both the optical and X-ray fluxes
together, by synchrotron radiation of secondary electrons.  Since the
internal absorption scenario requires a significant attenuation of the
VHE radiation over approximately two decades (see Figs.~\ref{fig:0229}
and \ref{fig:3c66amwl}), the secondary synchrotron component has to be
at least 4 decades broad (with additional broadening related to the
relativistic motion of the production region).  The strong magnetic
field required in the proton synchrotron model provides fast cooling
of the pairs, thus the radiation spectrum will be featureless, without
a cooling break.  In the case of a small radius of the production
region, the effective particle injection in the blob may be rather
narrow. But in this case the flux level would be significantly below
the observed flux.

\begin{deluxetable}{lccccccc}
  \tablecolumns{8} \tablewidth{0pc} \tablecaption{ The combination of the parameters used for the calculations of the model curves in     Figs.~\ref{fig:0229},~\ref{fig:3c66a} and~\ref{fig:3c66amwl} for the different EBL levels (first row): $p$ is the power-law index of the proton distribution, $B'$     is the magnetic field inside the blob, $T$ is the temperature of     the soft photon field, $\tau$ is the optical depth for the entire     source of soft photons in a region of radius $R$, $\tau_{\rm in}$ is the     optical depth inside the blob, $R_{\rm blob}'$ is the proper radius of     the blob, $\Gamma$ the bulk Lorentz factor, $\delta$ the Doppler     factor, $L_{\rm ph}$ the luminosity of the soft photon source, $L'_{\gamma}$ is the intrinsic luminosity of the $\gamma$-ray     source before Doppler boosting and internal absorption.  }  \tablehead{ \colhead{} &     \multicolumn{3}{c}{\textbf{1ES~0229+200}} & \colhead{} &
    \multicolumn{3}{c}{\textbf{3C~66A}} \\
    \cline{2-4} \cline{6-8} \\
    \colhead{Parameter} & \colhead{Fit 1} & \colhead{Fit 2} &     \colhead{Fit 3} & \colhead{} & \colhead{Fit 4a} & \colhead{Fit 4b}     & \colhead{Fit 5}} \startdata
  ${\rm EBL}$ & ${\rm F}1.6$ & ${\rm F}1.0$ & ${\rm F}1.6$ && ${\rm F}1.6$ &  ${\rm F}1.6$ & ${\rm F}1.0$ \\
  $p$                 & -0.5 &  -0.5 &  -0.5 && 2.2 & 2.2 & 2  \\
  $B'\,({\rm G})$           & 80 & 40 & 100 && 100 & 100 & 1.2\\
  $T\,({\rm K})$ & $7\times 10^3$ & $5\times 10^3$ & $10^5$ &&   $8\times 10^4$ & $8\times 10^4$ & $5\times 10^4$ \\
  $\tau$             & 3 & 3 & 5& & 2 & 2 & 1.6 \\
  $\tau_{\rm in}$ & 0.9 & 0.8 & 1.2& & 0.6 & $6\times 10^{-3}$ &   0.26 \\
  $R_{blob}'\,({\rm cm})$ & $10^{15}$ & $5\times 10^{15}$ & $5\times   10^{15}$ && $5\times 10^{17}$ & $5\times 10^{15}$ & $10^{18}$ \\
  $R\, ({\rm cm})$ & $3\times 10^{16}$ & $6\times 10^{17}$ &   $2.1\times 10^{17}$& & $7\times 10^{19}$ & $7\times 10^{19}$ &   $6\times 10^{19}$ \\
  $\Gamma$           & 10 & 30 & 10& & 40 & 40 & 10 \\
  $\delta$           & 11 & 8 & 8 & &15 & 15 & 4 \\
  $L_{ph}\,({\rm erg\,s^{-1}})$ & $2.3\times 10^{41}$ & $2.8\times 10^{42}$   & $3\times 10^{43}$& & $4\times 10^{45}$ & $4\times 10^{45}$ &   $2\times   10^{45}$ \\
  $L'_{\gamma}\,({\rm erg\,s^{-1}})$ & $5\times 10^{40}$ & $1.6\times   10^{41}$ & $2.9\times 10^{41}$ && $7\times 10^{42}$ & $7\times   10^{42}$ &   $4\times 10^{44}$ \\
\enddata
\label{table:parameters}
\end{deluxetable}

\section{Summary}

One of the most challenging issues of the physics of TeV blazars is
the rather hard intrinsic $\gamma$-ray spectra of some
representatives of this high-energy source population. Actually, the
reported spectra themselves are steep, with photon indexes
$\Gamma_{\rm int} \geq 3$.  However, the spectra of VHE $\gamma$-rays
after being corrected for the energy-dependent intergalactic
absorption become very hard, in some cases as hard as $\Gamma_{\rm
  int}=1.5$, assuming the EBL flux from \citet{franceschini08}. Note
however that some other recent models, in particular
\citet{dominguez10}, give very similar results.  Explanation of such
spectra faces serious difficulties within the standard blazar models.
Moreover, even a slight increase of the EBL flux at optical and near
IR wavelengths compared to the benchmark models (which, given the
significant uncertainties in the derivation of the EBL fluxes, cannot
be excluded) results in unusually hard intrinsic spectra with
$\Gamma_{\rm int} \leq 1$.

In this paper we studied the applicability of the idea of internal
absorption of $\gamma$-rays produced by highly magnetized blobs as a
result of synchrotron radiation of protons. While the main aspects of
the model have been developed and discussed in our previous paper
\citep{aharonian08}, in this work we tried to understand whether the
model can be applied to specific objects.  For this purpose, we have
chosen two  "difficult" representatives of this source
population,  1ES~0229+200 and 3C~66A, and allowed the EBL flux
to be somewhat higher than the lowest possible fluxes.  In particular, for the
EBL level consistent with the lower limit around $3.6\, \rm \mu m$
claimed by \cite{levenson08}, the intergalactic $\gamma$-$\gamma$
de-absorption of the VHE  flux detected from 3C~66A during
a flaring episode results in an extremely hard spectrum, with photon
index $\Gamma_{\rm int}=0.2$.  While such a hard spectrum cannot fit
into any existing VHE $\gamma$-ray production model, the scenario of
internal absorption of $\gamma$-rays produced via synchrotron radiation
of protons provides a reasonable explanation of both the hard TeV
spectrum and the high GeV flux observed during the flare.

In the case of 1ES~0229+200, the internal absorption scenario calls
for an extremely hard proton distribution, which can be provided for
example by the ``converter mechanism" \citep{derishev03}.  The
synchrotron radiation of secondary electron-positron pairs produced
inside the blob results in an additional (Doppler boosted) radiation
component, which can provide a self-consistent interpretation of the
non-thermal X-ray emission in both objects. 
However, since in the framework of the internal absorption scenario 
the secondary synchrotron component is expected to be quite broad and 
featureless, we failed to find a set of parameters which explains
simultaneously both the optical and X-ray data obtained from 3C66A.

The intrinsic absorption scenario allows a natural explanation of the
very hard intrinsic TeV spectra at the cost of a large attenuation of
the $\gamma$-ray flux around 100~GeV.  Quite remarkably, despite the
significant attenuation, this scenario does not enhance too dramatically
the required energy budget.  Indeed, since the correction for
intergalactic absorption requires the initial $\gamma$-ray spectra to
be very hard to begin with, with a photon index $\le 2$, the energy
requirement to reproduce such a spectrum is determined by the highest
energy part of the spectrum, which is not affected by internal
absorption.  Typically, the enhancement of the energy budget
introduced by internal absorption does not exceed a factor of 5, which
can be easily compensated by a slightly enhanced Doppler boosting of
the radiation.  Therefore the required intrinsic $\gamma$-ray
luminosities remain quite modest (see
Table~\ref{table:parameters}). Given the very high (almost 100\%)
efficiency of conversion of the proton energy to $\gamma$-rays through
the synchrotron radiation in the magnetized blob, the suggested
scenario can be treated as quite effective.

It is apparent that the interpretation of very hard $\gamma$-ray spectra by the internal
absorption depends on the "right" choice of several model parameters,
especially if we want to explain  the X-ray data by synchrotron radiation
of secondary electrons. Nevertheless, we should note that  the level of
hardening of the VHE $\gamma$-ray spectrum depends, in fact, only  on the
temperature of the photon field and the optical depth.  On the other
hand, while  the flux ratio of the VHE $\gamma$- and X-ray components
depends mainly on the  size of the production area, the location of the
secondary synchrotron peak is  sensitive to the strength of the magnetic
field, to the photon temperature and the bulk Lorentz factor. Finally, 
we note that certain radiation features of the scenario do not depend 
on the model parameters at all. This concerns, in particular, the slope 
of the $\gamma$-ray spectrum at GeV energies, and the shape of the 
X-ray spectrum.

The dependence of the results of radiation properties on
several parameters limits, to a certain extent, the predictive power
of the suggested model.  This is a consequence of complex
environment in blazars where several radiation and absorption
processes can proceed simultaneously. In this regard, the often used
one-zone models with consideration of only synchrotron and IC
radiation components produced in the same region, are quite useful
for understanding the basic aspects of the problem, but can hardly
properly describe the complex scenarios that take place in
blazars. In particular, the results of this paper demonstrate that
the internal absorption not only cannot be {\it a priory} excluded
from the consideration, but, in fact, in some cases can be invoked
for better explanation of observations of TeV $\gamma$-ray blazars.

\begin{acknowledgements}
The authors would like to thank Dr. L.~C.~Reyes for making available the VERITAS data of 3C~66A flare and Prof.~F.~Tavecchio for the SWIFT data of 1ES~0229+200.
\end{acknowledgements}


\end{document}